# Title: Every-other-layer Dipolar Excitons in a Spin-Valley locked Superlattice

Authors: Yinong Zhang[1], Chengxin Xiao[2,3], Dmitry Ovchinnikov[1], Jiayi Zhu[1], Xi Wang[1,4], Takashi Taniguchi[5], Kenji Watanabe[6], Jiaqiang Yan[7], Wang Yao[2,3,*], and Xiaodong Xu[1,8,*]

[1]Department of Physics, University of Washington, Seattle, WA, USA
[2]Department of Physics, University of Hong Kong, Hong Kong, China
[3]HKU-UCAS Joint Institute of Theoretical and Computational Physics at Hong Kong, China
[4]Department of Chemistry, University of Washington, Seattle, WA, USA
[5]International Center for Materials Nanoarchitectonics, National Institute for Materials Science, Tsukuba, Ibaraki 305-0044, Japan
[6]Research Center for Functional Materials, National Institute for Materials Science, Tsukuba, Ibaraki 305-0044, Japan
[7]Materials Science and Technology Division, Oak Ridge National Laboratory, Oak Ridge, Tennessee, 37831, USA
[8]Department of Materials Science and Engineering, University of Washington, Seattle, WA, USA

Correspondence to: wangyao@hku.hk; xuxd@uw.edu

**Abstract: Monolayer semiconducting transition metal dichalcogenides possess broken inversion symmetry and strong spin-orbit coupling, which leads to unique spin-valley locking effect[1-6]. In 2H stacked pristine multilayers, the spin-valley locking yields an electronic superlattice structure, where alternating layers correspond to barrier and quantum well respectively, conditioned on the spin-valley indices[2,7-11]. Here, we show that the spin-valley locked superlattice hosts a new kind of dipolar excitons with the electron and hole constituents separated in an every-other-layer configuration, i.e., either in two even or two odd layers. Such excitons become optically bright via hybridization with intralayer excitons, displaying multiple anti-crossing patterns in optical reflection spectrum as the dipolar exciton is tuned through the intralayer resonance by electric field. The reflectance spectra also reveal an excited state orbital of the every-other-layer exciton, pointing to a sizable binding energy in the same order of magnitude as the intralayer exciton. As layer thickness increases, the dipolar exciton can form one-dimensional Bose-Hubbard chain displaying a layer number dependent fine-structures in the reflectance spectra. Our work reveals a distinct valleytronic superlattice with highly tunable dipolar excitons for exploring light-matter interactions[12-15].**

**Main text**

In van der Waals crystals, layer stacking arrangement plays a critical role in determining the material physical properties. Examples include the localized charge hopping and thus flat band formation in magic angle twisted graphene[16,17], the sign switching of interlayer magnetic coupling between rhombohedral and monoclinic stacked 2D magnet $CrI_3$[18,19], and the formation of interfacial ferroelectricity in AB/BA stacked hexagonal boron nitride (hBN)[20-22]. For transition metal dichalcogenides, such as $WSe_2$, its monolayer feature a sizable spin splitting at the band edges located at the $\pm K$ valleys at the corners of hexagonal Brillouin zone, where time reversal dictates a valley-contrasted sign of the splitting[1]. Multilayers of the 2H stacking feature the 180° rotation between adjacent layers, which is an operation that swaps the spin states between the valleys (Fig. 1a). Consequently, for carrier of a given spin-valley index, the dispersion edge alternates between the two split values in even and odd layers. These realize perfect superlattice confinements in the out-of-plane direction (Fig. 1b): for a spin up carrier in the K valley $(K, \uparrow)$, the even layers serve as energy barriers defining the odd layers as quantum wells (QW), while for carrier with opposite spin-valley index $(K, \downarrow)$ the odd and even layers swap their roles.

In such spin-valley configured superlattice, the barrier height corresponds to the spin splitting[11]. Use $WSe_2$ as an example, the spin-orbital coupling induced spin splitting is as large as ~ 480 meV in the valence bands[23,24]. Therefore, the low energy holes are well confined in the QW regions (Fig. 1b). Electrons see a lower barrier (~30 meV), but their interlayer hopping is forbidden at the $\pm K$ points by the rotational symmetry (Fig. 1c). These are reminiscent of the semiconductor superlattices from the periodical growth of two alternating III-V compounds which underlies a broad range of important optical phenomena and applications, but now realized in a homogeneous compound, featuring spin-valley dependent confinement (Fig. 1b).

In this work, we report observation of a new type of dipolar excitons in such spin-valley locked superlattices in multilayer $WSe_2$. The dipolar exciton, with electron and hole separated in an every-other-layer configuration, has large electric dipole but vanishing optical dipole (Fig. 1d). However, it can become optically bright by hybridizing with the intralayer excitons via a weak next-nearest-neighbor interlayer hopping of carriers conserving spin and valley indices (Figs. 1c&d). We first present the results from a trilayer $WSe_2$ as the building block for such superlattice, i.e., a double QWs.

Figure 1e shows the differential reflectance spectrum normalized by the background reflectance (dR/R) in trilayer $WSe_2$ without applying gate voltage. The inset of Fig. 1e is the optical microscope image of the device, where the $WSe_2$ is sandwiched by hBN in a dual-gated geometry (see Methods). The doping dependence of dR/R at zero electric field is shown in Fig. 1f. There are three resonances at charge neutrality, while the broad background at high energy is caused by optical interference. The spectral features near 1.705 eV and 1.798 eV correspond to the 1S and 2S intralayer excitons, respectively[25-27]. The 1S intralayer exciton state is schematically indicated in Fig. 1d by the electron-hole pair enclosed by orange ellipse. Besides, there's a relatively weaker state at 1.767eV, which is not seen from monolayer and bilayer $WSe_2$ spectra (see Extended Data Fig. 1). The origin of this feature, corresponding to the every-other-layer dipolar exciton (DX) with the electron-hole pair enclosed by blue ellipse shaded in Fig. 1d, will become clear later in the electrical field dependent measurements. The larger spatial separation of the electron and hole pair leads to a smaller binding energy of the dipolar exciton compared to the intralayer 1S exciton, and thus higher energy in optical spectrum.

We examine optical reflection versus the out-of-plane electric field $E_z$ at charge neutrality (Fig. 2a. The photoluminescence measurement is shown in Extended Data Fig. 2). As expected, the peak position of both 1S and 2S intralayer excitons are independent of $E_z$. In contrast, the exciton feature at 1.767 eV evolves into two branches, whose energies shift linearly with $E_z$ and develop an X pattern centering at zero-field resonance. Since there are two energy degenerate dipolar excitons but with opposite electric dipoles at $E_z = 0$, the applied electric field breaks their degeneracy, linearly decreases (increases) the exciton energy, and forms the observed X pattern. By fitting the stark effect, we can determine the electric dipole to be $d = \frac{\varepsilon_{TMD}}{\varepsilon_{hBN}} \frac{\Delta Energy}{\Delta Ez} \sim 1.4\, e \cdot nm$, using dielectric constant $\varepsilon_{TMD} \sim 7$ and $\varepsilon_{hBN} \sim 3$. This extracted dipole is exactly twice of the usual interlayer exciton observed in heterobilayer TMDs, demonstrating that the electron and hole are now separated in every other layer.

Figure 2a also shows several avoided-crossing features as the dipolar exciton is tuned across the intralayer exciton resonances[28]. For instances, as $E_z$ is around ±0.1 V/nm (±0.06 V/nm), the energy of the dipolar exciton is nearly resonant with intralayer 1S (2S) excitons, where dipolar excitons hybridize with intralayer ones. This hybridization stems from the carrier hopping between layer 1 and 3 (Figs. 1c&d), and is crucial for the observation of the every-other-layer DX. We note that such DX has vanishing optical dipole by itself, and becomes visible only through hybridization with bright intralayer exciton states.

To quantitatively understand the observation, we examine the coupling of the relevant exciton states: intralayer 1S, 2S excitons in layer 1 and 3, and the two dipolar excitons with opposite electric dipoles. The Hamiltonian is written as

$$H = \begin{pmatrix} \varepsilon_{1S} & 0 & t_1 & t_1 & 0 & 0 \\ 0 & \varepsilon_{1S} & t_1 & t_1 & 0 & 0 \\ t_1 & t_1 & \varepsilon_{DX} - E_z \cdot d & 0 & t_2 & t_2 \\ t_1 & t_1 & 0 & \varepsilon_{DX} + E_z \cdot d & t_2 & t_2 \\ 0 & 0 & t_2 & t_2 & \varepsilon_{2S} & 0 \\ 0 & 0 & t_2 & t_2 & 0 & \varepsilon_{2S} \end{pmatrix}$$

where, $\varepsilon_{1S}$, $\varepsilon_{2S}$, $\varepsilon_{DX}$ denote the energies of intralayer and dipolar excitons at zero electrical field. Term $\pm E_z \cdot d$ corresponds to the Stark shift of two dipolar excitons DX$_{\uparrow\downarrow}$. The coupling between the dipolar exciton and 1S (2S) intralayer exciton is through the carrier hopping with amplitude $t_1$ ($t_2$). By fitting the eigenvalues versus $E_z$ to Fig. 2a, we estimate $t_1 = t_2 \sim 5.5$ meV. After parametrizing the linewidth and emission intensity of the intralayer and dipolar excitons (see Extended Data Fig. 3 and Methods), the calculation well reproduces the differential reflectance spectra in Fig. 2b, which further supports our interpretation of the results.

We take the derivative of dR/R spectra in Fig. 2a versus the energy axis, $d(dR/R)/d\varepsilon$. As shown in Fig. 2c, it reveals subtle spectral features. A second X pattern above the intralayer 2S state appears, centering at 1.813 eV. This X pattern exhibits the stark effect with the same slope corresponding to electric dipole $d \sim 1.4\, e \cdot nm$. Therefore, we attribute this second anti-crossing to the higher orbital state of the every-other-layer exciton, which becomes visible through hybridizing with the intralayer 2S state. The energy separation between the ground and excited states of this every-other-layer DX is about 46 meV at $E_z = 0$, which sets the lower bound of the binding energy. In comparison to the intralayer 1S-2S exciton energy difference of about 93 meV, the binding energy of the DX is of the same order of magnitude. The observation of such a large energy scale for the every-other-layer binding is remarkable, which suggests the screening effect in trilayer is still strongly suppressed.

Every-other-layer dipolar exciton relies strongly on the spin-valley locked spin superlattice chain. If we break this periodicity in the stacking order, the dipolar exciton can be switched off. For this purpose, we measured artificially twisted double bilayer $WSe_2$ samples. Figures 3a-b show the dR/R intensity plot of a naturally 2H-stacked 4-layer (Figs. 3a), near 0° twist (Fig. 3b) double bilayers, and near 60 twisted double bilayers (Fig. 3c). In the natural 4-layer sample, the X spectral pattern vs $E_z$ demonstrates the every-other-layer dipolar exciton, which is formed between the layer number with same parity, i.e., either $1^{st}/3^{rd}$ layer or $2^{nd}/4^{th}$ layer. Similar X pattern is observed in the zero-degree stacked double bilayers. In contrast, the 60-degree stacked double bilayers shows drastically different spectra features compared to both naturally 2H and zero-degree twisted samples. As shown in Fig. 3c, only the 1S intralayer exciton is observed, which is insensitive to $E_z$.

The 0° twisted sample resembles the natural 4-layer $WSe_2$ in their ABAB stacking order, where each layer sits on top of another with 180° rotation (Fig. 3d). Due to the small twist angle between the double bilayers, moiré superlattices form at the interface between the two $WSe_2$ bilayers[29-31]. In Fig. 3b, the intralayer 1S (1.706 eV), 2S (1.803 eV) and dipolar exciton (1.763 eV) states are all observed as well as the anti-crossing features near the intralayer 1S resonance. An electric dipole of $1.4\ e \cdot nm$ is obtained, evidence for the every-other-layer configuration of the dipolar exciton. Features from 0° twist sample exhibits broader linewidth than natural 4-layer sample, possibly due to the inhomogeneity of 2H domain structures formed at the twisted interface. In comparison, the every-other-layer dipolar exciton is absent from the 60° twisted double bilayer $WSe_2$ (Fig. 3c). Different from 0° twisted and natural 4-layer samples, 60° twisted sample is stacked in an ABBA order (Fig. 3d). In such structures, carrier hopping between the two middle layers are allowed, and every-other-layer DX is no longer favored as a low energy configuration.

Finally, we examine dipolar excitons in samples with different thickness. We do not observe exciton feature with such a large electrical dipole in either monolayer or bilayer samples (see Extended Data Fig. 1). This confirms the basic unit for such an exciton is trilayer. While in samples thicker than 4 layers, fine structures of multiple X patterns appear in the differential reflectance spectra. In 5-layer sample (Fig. 4a), there are two X patterns with a modest energy splitting of about 20 meV. Possible mechanisms to split the X pattern include a second order carrier hopping process which couples two dipolar excitons through an intralayer exciton as the virtual intermediate state. We use $DX_{ij}$ to denote the dipolar exciton between the $i^{th}$ and $j^{th}$ layer. As schematically shown in Fig. 4c, through hole hopping process between the $1^{st}$ and $3^{rd}$ layer, the dipolar exciton $D_{13}$ is firstly coupled to the intralayer state at the $3^{rd}$ layer, which is then coupled to another dipole-up exciton $DX_{35}$ through an electron hopping process between the $3^{rd}$ and $5^{th}$ layer. However, given the carrier hopping amplitude $t \sim 5.5$ meV we previously derive from fitting the trilayer $WSe_2$'s spectra, the above second order coupling process between two dipolar excitons is expected to give a smaller splitting than what we observed.

A more plausible explanation is the inhomogeneous dielectric environment along the vertical axis of the device stacking[32]. The large layer separation of the every-other-layer dipolar exciton leads to a larger Bohr radius, making it more sensitive to the surrounding dielectric environment. For example, comparing the $DX_{13}$ and $DX_{24}$ in a five-layer (Fig. 4d), the former is influenced more by the hBN layers which has different dielectric screening from the TMD layers. Such different dielectric environment for different dipolar excitons results in slight difference in their binding energies. Likewise, in 7-layer sample, $DX_{13}$ ($DX_{57}$), $DX_{24}$ ($DX_{46}$) and $DX_{35}$ will split into three different branches for the same reason, which agrees with what we observed in Fig. 4b. In comparison, in 4-layer WSe$_2$, $DX_{13}$ and $DX_{24}$ do not couple to each other, and have the same dielectric environment under the symmetric geometry with hBN capping layer and substrate. The two thus have similar resonance energy (Fig. 3a). These every-other-layer excitons with the large electric dipoles in the spin-valley locked superlattice imply the possibilities for studying and engineering many-body effects of interacting bosons, as well as potential in nanophotonics and quantum optoelectronics by controlling the excitonic emission through layer-by-layer engineering.

**Methods**

**Device fabrication.** Flakes of hexagonal Boron Nitride (hBN), graphite and WSe$_2$ are mechanically exfoliated from bulk crystals on Si/SiO$_2$ substrates. They are selected using optical microscope and atomic force microscope (AFM). The WSe$_2$ layer number is determined by AFM and color contrast, assisted by second-harmonic generation. The heterostructure is assembled layer-by-layer using a dry transfer technique. The twisted double bilayer WSe$_2$ is fabricated using a 'tear-and-stack' technique, during which part of the bilayer WSe2 large flake is torn and picked up and stacked onto the remaining part with a twist angle which is precisely controlled by a rotary stage. Finally, the Cr/Au contact are patterned with standard electron beam lithography and evaporation.

**Optical measurements.** The differential reflectance measurement is performed using a home-built confocal microscope in reflection geometry. The sample is mounted in an exchange-gas cooled cryostat with temperatures down to 5K unless otherwise specified. A tungsten halogen white light source is used to excite the sample through a single mode optical fiber. The beam diameter on the sample is about 1 μm. The reflected light passes through a 50 μm pin hole and is collected by a spectrometer with a silicon charge-coupled device. Polarization resolved measurements are performed with a set of broad-band half-wave plates, quarter wave plates and linear polarizers. The differential reflection (dR/R) spectrum is obtained as $dR/R \equiv (R' - R)/R$, by comparing the reflected light intensity from the sample (R') to that of a featureless spectrum (R) from graphite/hBN/graphite area in the same sample.

**Calibration of doping density and electric field.** The doping densities in the heterobilayer are determined from the applied gate voltages based on a parallel-plate capacitor model[33]. The thickness of hBN is determined by atomic force microscopy, typically around 10-20nm thick. The doping density is calculated as $C_b \Delta V_b + C_t \Delta V_t$, where $C_t$ and $C_b$ are the capacitances of the top and bottom gates and $\Delta V_t$ and $\Delta V_b$ are the applied gate voltages relative to the level of the valence/conduction band edge. The geometric capacitance is calculated using $C_{t,b} = \varepsilon_0 \varepsilon_{BN}/d_{hBN}$ with the dielectric constant of hBN $\varepsilon_{hBN} \sim 3$ (Ref[34]). The out-of-plane electric displacement field is calculated using $D = (C_b \Delta V_b - C_t \Delta V_t)/2\varepsilon_0$ and the electric field is calculated using $E_Z = D/\varepsilon_{hBN}$.

**Modeling and simulation:** In trilayer, the relevant excitonic basis states for the observed hybridization are: $|1S^{top}\rangle$, $|1S^{bottom}\rangle$, $|2S^{top}\rangle$, $|2S^{bottom}\rangle$, $|DX_\uparrow\rangle$, $|DX_\downarrow\rangle$, denoting respectively the 1S and 2S orbital intralayer excitons in top and bottom layer, and every-other-layer dipolar excitons with electrical dipole pointing up and down. The diagonal part of the Hamiltonian can be written as:

$$H_0 = \begin{pmatrix} \varepsilon_{1S} & 0 & 0 & 0 & 0 & 0 \\ 0 & \varepsilon_{1S} & 0 & 0 & 0 & 0 \\ 0 & 0 & \varepsilon_{DX} - E_z \cdot d & 0 & 0 & 0 \\ 0 & 0 & 0 & \varepsilon_{DX} + E_z \cdot d & 0 & 0 \\ 0 & 0 & 0 & 0 & \varepsilon_{2S} & 0 \\ 0 & 0 & 0 & 0 & 0 & \varepsilon_{2S} \end{pmatrix} \quad (1a)$$

Where $\pm E_z \cdot d$ corresponds to the Stark shift of $|DX_{\downarrow\uparrow}\rangle$ away from their zero-field energy $\varepsilon_{DX}$. Interlayer hopping of electron/hole leads to an off-diagonal coupling $H_1$ between $DX_{\downarrow\uparrow}$ and intralayer states (both 1S and 2S):

$$H_1 = \begin{pmatrix} 0 & 0 & t_1 & t_1 & 0 & 0 \\ 0 & 0 & t_1 & t_1 & 0 & 0 \\ t_1 & t_1 & 0 & 0 & t_2 & t_2 \\ t_1 & t_1 & 0 & 0 & t_2 & t_2 \\ 0 & 0 & t_2 & t_2 & 0 & 0 \\ 0 & 0 & t_2 & t_2 & 0 & 0 \end{pmatrix} \quad (1b)$$

where $t_1$ ($t_2$) stands for the coupling strength between $DX_{\downarrow\uparrow}$ and intralayer 1S (2S) state.

After diagonalization of $H = H_0 + H_1$ by $H_D = U^\dagger H U$, we obtain six eigenstates ($|X_1\rangle$, $|X_2\rangle$, $|X_3\rangle$, $|X_4\rangle$, $|X_5\rangle$, $|X_6\rangle$),

$$\begin{pmatrix} |X_1\rangle \\ |X_2\rangle \\ |X_3\rangle \\ |X_4\rangle \\ |X_5\rangle \\ |X_6\rangle \end{pmatrix} = U^\dagger \begin{pmatrix} |1S^{top}\rangle \\ |1S^{bottom}\rangle \\ |DX_\uparrow\rangle \\ |DX_\downarrow\rangle \\ |2S^{top}\rangle \\ |2S^{bottom}\rangle \end{pmatrix}$$

and eigenenergies $H_D = diag(\varepsilon_1(E_z), \varepsilon_2(E_z), \varepsilon_3(E_z), \varepsilon_4(E_z), \varepsilon_5(E_z), \varepsilon_6(E_z))$, as function of electric field. These mixed eigenstates are close to the base vectors, except around $\varepsilon_{DX} \pm E_z \cdot d \approx \varepsilon_{1S/2S}$ regions, where $DX_{\downarrow\uparrow}$ hybridize with intralayer 1S/2S and anti-crossing features appear. By fitting the resonance features in the measured dR/R data (red curves in Extended Data Fig. 3a), we obtain the coupling strength $t_1 \approx t_2 \approx 5.5 \ meV$.

We further model the spectral density of each excitonic state as a Gaussian distribution with a bandwidth $w_i$ and intensity $I_i$ ($i = 1S^{top/bottom}$, $DX_{\downarrow/\uparrow}$, $2S^{top/bottom}$), the optical reflectance spectrum as a function of incidence energy $\varepsilon$ and electric field $E_z$ is given by:

$$I(\varepsilon, E_z) = \sum_j \sum_i I_i |U_{ij}|^2 e^{-(\varepsilon - \varepsilon_j(E_z))^2 / 2w_i^2}$$

Extended Data Fig. 3b is a plot using $I_{1S}^{top} = I_{1S}^{bottom} = 0.5$, $I_{DX,\downarrow/\uparrow} = 0$, $I_{2S}^{top} = I_{2S}^{bottom} = 0.1$, $w_{1S} = 5.5 \ meV$, $w_{2S} = 4.5 \ meV$, where we have assumed negligible optical oscillator strength of the every-other-layer DX. Nevertheless, $DX_{\downarrow\uparrow}$ states become visible through hybridization with intralayer 1S/2S states.

In 5-layer WSe$_2$, the corresponding Hamiltonian becomes:

$$H = H_0 + H_1 = \begin{pmatrix} H_{1S} & 0 & 0 \\ 0 & H_{DX} & 0 \\ 0 & 0 & H_{2S} \end{pmatrix} + \begin{pmatrix} 0 & H_{1S-DX} & 0 \\ H_{1S-DX}^\dagger & 0 & H_{2S-DX}^\dagger \\ 0 & H_{2S-DX} & 0 \end{pmatrix}$$

where

$$H_{1S/2S} = \begin{pmatrix} \varepsilon_{1S/2S}^{(1)} & 0 & 0 & 0 & 0 \\ 0 & \varepsilon_{1S/2S}^{(2)} & 0 & 0 & 0 \\ 0 & 0 & \varepsilon_{1S/2S}^{(3)} & 0 & 0 \\ 0 & 0 & 0 & \varepsilon_{1S/2S}^{(4)} & 0 \\ 0 & 0 & 0 & 0 & \varepsilon_{1S/2S}^{(5)} \end{pmatrix}$$

$\varepsilon_{1S/2S}^{(i)}$ standing for the energy of intralayer 1S/2S exciton within the i$^{th}$ layer, and

$$H_{DX} = \begin{pmatrix} \varepsilon_{DX}^{(13)} & 0 & 0 & 0 & 0 & 0 \\ 0 & \varepsilon_{DX}^{(31)} & 0 & 0 & 0 & 0 \\ 0 & 0 & \varepsilon_{DX}^{(35)} & 0 & 0 & 0 \\ 0 & 0 & 0 & \varepsilon_{DX}^{(53)} & 0 & 0 \\ 0 & 0 & 0 & 0 & \varepsilon_{DX}^{(24)} & 0 \\ 0 & 0 & 0 & 0 & 0 & \varepsilon_{DX}^{(42)} \end{pmatrix}$$

$\varepsilon_{DX}^{(ij)}$ standing for the energy of every-other-layer dipolar exciton between the i$^{th}$ and j$^{th}$ layer.

$H_{1S-DX}$ ($H_{2S-DX}$) stands for the off-diagonal coupling between 1S and DX (2S and DX) states through the electro or hole interlayer hopping.

$$H_{1S-DX/2S-DX} = \begin{pmatrix} t_{1/2} & t_{1/2} & 0 & 0 & 0 & 0 \\ 0 & 0 & 0 & 0 & t_{1/2} & t_{1/2} \\ t_{1/2} & t_{1/2} & t_{1/2} & t_{1/2} & 0 & 0 \\ 0 & 0 & 0 & 0 & t_{1/2} & t_{1/2} \\ 0 & 0 & t_{1/2} & t_{1/2} & 0 & 0 \end{pmatrix}$$

Using the same fitting parameters $t_1 = t_2 = 5.5\ meV$ as in the trilayer case, the eigenenergies of the five-layer Hamiltonian are plotted as the red curves in Extended Data Fig. 3c. Note that, considering the different dielectric environment of DX$_{13/35}$ and DX$_{24}$, we have assigned different values to $\varepsilon_{DX}^{(24/42)}$ and $\varepsilon_{DX}^{(13/31/35/53)}$. Extended Data Fig. 3d shows the optical spectrum, with the spectral intensity and width parameters $I_{1S}^{(i)} = 0.2$, $I_{DX,\downarrow/\uparrow}^{(ij)} = 0$, $I_{2S}^{(i)} = 0.06$, $w_{1S} = 5.5\ meV$, $w_{2S} = 4.5\ meV$.

**Acknowledgements:** This work is mainly supported by DoE BES under award DE-SC0018171. Sample fabrication and PFM characterization are partially supported by ARO MURI program (grant no. W911NF-18-1-0431). The AFM-related measurements were performed using instrumentation supported by the U.S. National Science Foundation through the UW Molecular

Engineering Materials Center (MEM·C), a Materials Research Science and Engineering Center (DMR-1719797). WY and CX acknowledge support by the University Grant Committee/Research Grants Council of Hong Kong SAR (AoE/P-701/20, HKU SRFS2122-7S05). WY also acknowledge support from Tencent Foundation. Bulk WSe$_2$ crystal growth and characterization by JY is supported by the US Department of Energy, Office of Science, Basic Energy Sciences, Materials Sciences and Engineering Division. K.W. and T.T. acknowledge support from the JSPS KAKENHI (Grant Numbers 19H05790, 20H00354 and 21H05233). XX acknowledges support from the State of Washington funded Clean Energy Institute and from the Boeing Distinguished Professorship in Physics.

**Author contributions:** XX and WY conceived the project. YZ fabricated samples and performed measurements, assisted by DO, JZ, XW. YZ, XX, CX, WY analyzed and interpreted the results. CX and WY performed calculations. TT and KW synthesized the hBN crystals. JY synthesized and characterized the bulk WSe$_2$ crystals. YZ, XX, CX, WY wrote the paper with input from all authors. All authors discussed the results.

**Competing interests:** The authors declare no competing financial interests.

**Data Availability:** The datasets generated during and/or analyzed during this study are available from the corresponding author upon reasonable request.

**References:**

1  Xiao, D., Liu, G.-B., Feng, W., Xu, X. & Yao, W. Coupled spin and valley physics in monolayers of MoS 2 and other group-VI dichalcogenides. *Physical review letters* **108**, 196802 (2012).
2  Jones, A. M. *et al.* Spin–layer locking effects in optical orientation of exciton spin in bilayer WSe2. *Nature Physics* **10**, 130-134 (2014).
3  Mak, K. F., He, K., Shan, J. & Heinz, T. F. Control of valley polarization in monolayer MoS2 by optical helicity. *Nature nanotechnology* **7**, 494-498 (2012).
4  Wu, S. *et al.* Electrical tuning of valley magnetic moment through symmetry control in bilayer MoS2. *Nature Physics* **9**, 149-153 (2013).
5  Zeng, H., Dai, J., Yao, W., Xiao, D. & Cui, X. Valley polarization in MoS2 monolayers by optical pumping. *Nature nanotechnology* **7**, 490-493 (2012).
6  Cao, T. *et al.* Valley-selective circular dichroism of monolayer molybdenum disulphide. *Nature communications* **3**, 1-5 (2012).
7  Gong, Z. *et al.* Magnetoelectric effects and valley-controlled spin quantum gates in transition metal dichalcogenide bilayers. *Nature communications* **4**, 1-6 (2013).
8  Shi, Q. *et al.* Bilayer WSe2 as a natural platform for interlayer exciton condensates in the strong coupling limit. *Nature Nanotechnology*, 1-6 (2022).
9  Yankowitz, M., McKenzie, D. & LeRoy, B. J. Local spectroscopic characterization of spin and layer polarization in WSe 2. *Physical review letters* **115**, 136803 (2015).
10  Bawden, L. *et al.* Spin–valley locking in the normal state of a transition-metal dichalcogenide superconductor. *Nature communications* **7**, 1-6 (2016).
11  Slobodeniuk, A. O. *et al.* Fine structure of K-excitons in multilayers of transition metal dichalcogenides. *2D Materials* **6**, 025026 (2019).

12    Wang, G. *et al.* Colloquium: Excitons in atomically thin transition metal dichalcogenides. *Reviews of Modern Physics* **90**, 021001 (2018).
13    Dufferwiel, S. *et al.* Exciton–polaritons in van der Waals heterostructures embedded in tunable microcavities. *Nature communications* **6**, 1-7 (2015).
14    Zhang, L., Gogna, R., Burg, W., Tutuc, E. & Deng, H. Photonic-crystal exciton-polaritons in monolayer semiconductors. *Nature communications* **9**, 1-8 (2018).
15    Liu, X. *et al.* Strong light–matter coupling in two-dimensional atomic crystals. *Nature Photonics* **9**, 30-34 (2015).
16    Bistritzer, R. & MacDonald, A. H. Moiré bands in twisted double-layer graphene. *Proceedings of the National Academy of Sciences* **108**, 12233-12237 (2011).
17    Cao, Y. *et al.* Unconventional superconductivity in magic-angle graphene superlattices. *Nature* **556**, 43-50 (2018).
18    Song, T. *et al.* Switching 2D magnetic states via pressure tuning of layer stacking. *Nature materials* **18**, 1298-1302 (2019).
19    Li, T. *et al.* Pressure-controlled interlayer magnetism in atomically thin $CrI_3$. *Nature materials* **18**, 1303-1308 (2019).
20    Yasuda, K., Wang, X., Watanabe, K., Taniguchi, T. & Jarillo-Herrero, P. Stacking-engineered ferroelectricity in bilayer boron nitride. *Science* **372**, 1458-1462 (2021).
21    Woods, C. *et al.* Charge-polarized interfacial superlattices in marginally twisted hexagonal boron nitride. *Nature communications* **12**, 1-7 (2021).
22    Vizner Stern, M. *et al.* Interfacial ferroelectricity by van der Waals sliding. *Science* **372**, 1462-1466 (2021).
23    Liu, G.-B., Shan, W.-Y., Yao, Y., Yao, W. & Xiao, D. Three-band tight-binding model for monolayers of group-VIB transition metal dichalcogenides. *Physical Review B* **88**, 085433 (2013).
24    Wilson, N. R. *et al.* Determination of band offsets, hybridization, and exciton binding in 2D semiconductor heterostructures. *Science advances* **3**, e1601832 (2017).
25    Król, M. *et al.* Exciton-polaritons in multilayer $WSe_2$ in a planar microcavity. *2D Materials* **7**, 015006 (2019).
26    Arora, A. *et al.* Valley-contrasting optics of interlayer excitons in Mo-and W-based bulk transition metal dichalcogenides. *Nanoscale* **10**, 15571-15577 (2018).
27    Raiber, S. *et al.* Ultrafast pseudospin quantum beats in multilayer $WSe_2$ and $MoSe_2$. *Nature Communications* **13**, 4997, doi:10.1038/s41467-022-32534-3 (2022).
28    Shimazaki, Y. *et al.* Strongly correlated electrons and hybrid excitons in a moiré heterostructure. *Nature* **580**, 472-477 (2020).
29    Yoo, H. *et al.* Atomic and electronic reconstruction at the van der Waals interface in twisted bilayer graphene. *Nature materials* **18**, 448-453 (2019).
30    Zhang, C. *et al.* Interlayer couplings, Moiré patterns, and 2D electronic superlattices in $MoS_2/WSe_2$ hetero-bilayers. *Science advances* **3**, e1601459 (2017).
31    McGilly, L. J. *et al.* Visualization of moiré superlattices. *Nature Nanotechnology* **15**, 580-584 (2020).
32    Raja, A. *et al.* Coulomb engineering of the bandgap and excitons in two-dimensional materials. *Nature communications* **8**, 1-7 (2017).
33    Wang, Z., Zhao, L., Mak, K. F. & Shan, J. Probing the spin-polarized electronic band structure in monolayer transition metal dichalcogenides by optical spectroscopy. *Nano letters* **17**, 740-746 (2017).

34      Movva, H. C. *et al.* Density-dependent quantum Hall states and Zeeman splitting in monolayer and bilayer WSe 2. *Physical review letters* **118**, 247701 (2017).

**Figures**

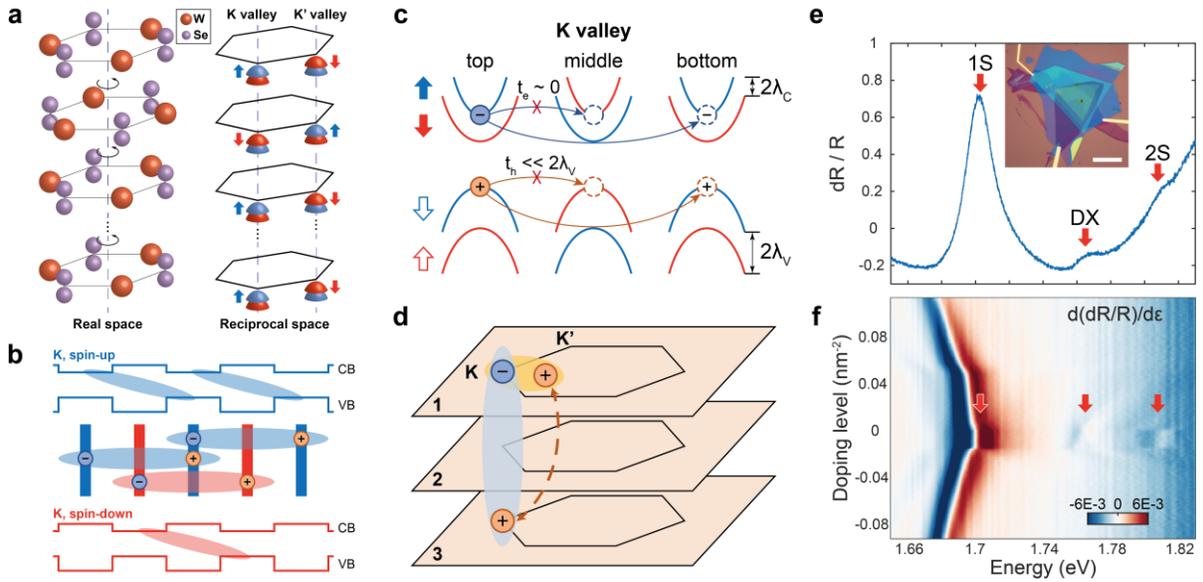

**Figure 1 | Spin-valley locked superlattice and every-other-layer exciton. a,** Crystal structure of multilayer 2H-WSe$_2$ in real space (left) and spin-valley coupled band edges in reciprocal space (right). Blue and red arrows stand for spin up and down. **b,** Schematic of the spin-valley dependent superlattice confinement of band edge carriers in 2H-WSe$_2$ multilayer, and every-other-layer dipolar excitons (DXs). CB and VB denote the conduction and valence band. Light blue/red ellipses represent DXs formed in odd/even layers. **c,** Interlayer hopping between adjacent layers is forbidden due to spin-valley locking, while every-other-layer hopping is allowed. **d,** Coupling between DX (light blue) and intralayer (orange) excitons in trilayer geometry. **e,** Differential optical reflection spectrum (dR/R) from trilayer WSe$_2$ at charge neutrality and zero field. DX and intralayer 1S, 2S excitons are indicated by red arrows. Inset: Optical microscope image of the dual-gated device (scale bar = 20 μm). **f,** Doping dependence of differentiated dR/R from trilayer WSe$_2$ at zero electric field.

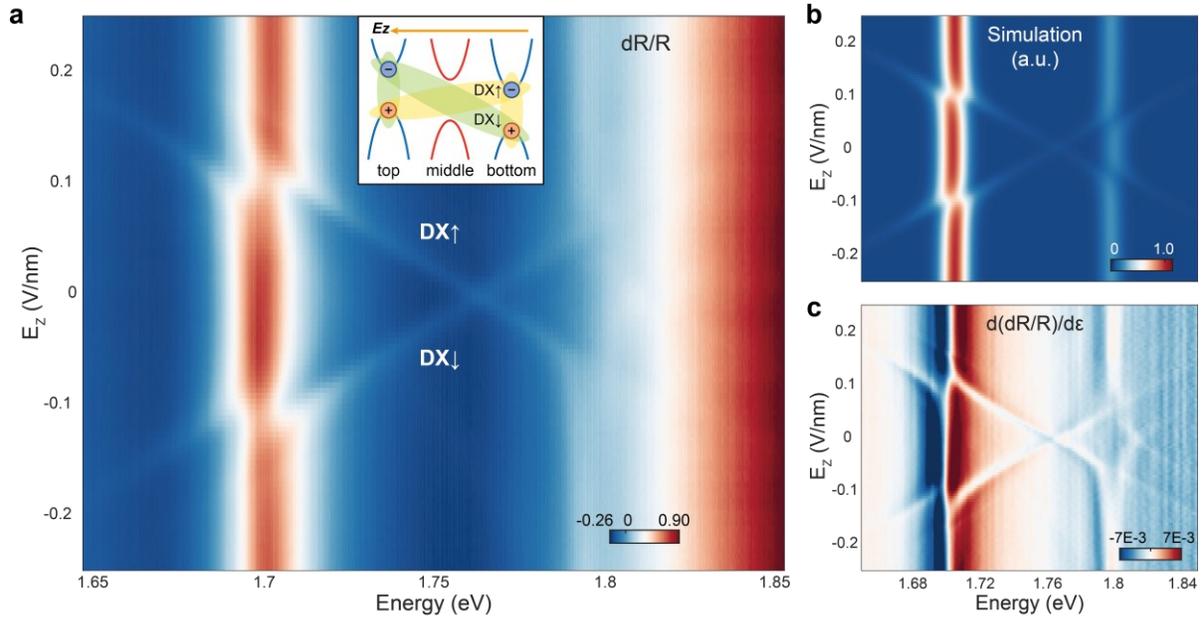

**Figure 2 | Electric field tuning of exciton hybridization. a,** Electric field dependence of dR/R from trilayer WSe$_2$ at charge neutrality. Anti-crossing feature signifies the hybridization between DX and intralayer 1S/2S excitons. DX$_\uparrow$ (DX$_\downarrow$) stands for dipolar exciton with hole and electron in the top (bottom) and bottom (top) layers of the trilayer, respectively. Inset: Schematic of energy bands and exciton structures under positive electric field, where DX$_\uparrow$ can hybridize with the intralayer 1S exciton close by in energy. **b,** Calculated electric field dependent spectral density. **c,** Reflectance spectrum differentiated with respective to photon energy (d(dR/R)/dε), where an excited state orbital of DX$_{\uparrow\downarrow}$ becomes visible as a higher energy X pattern.

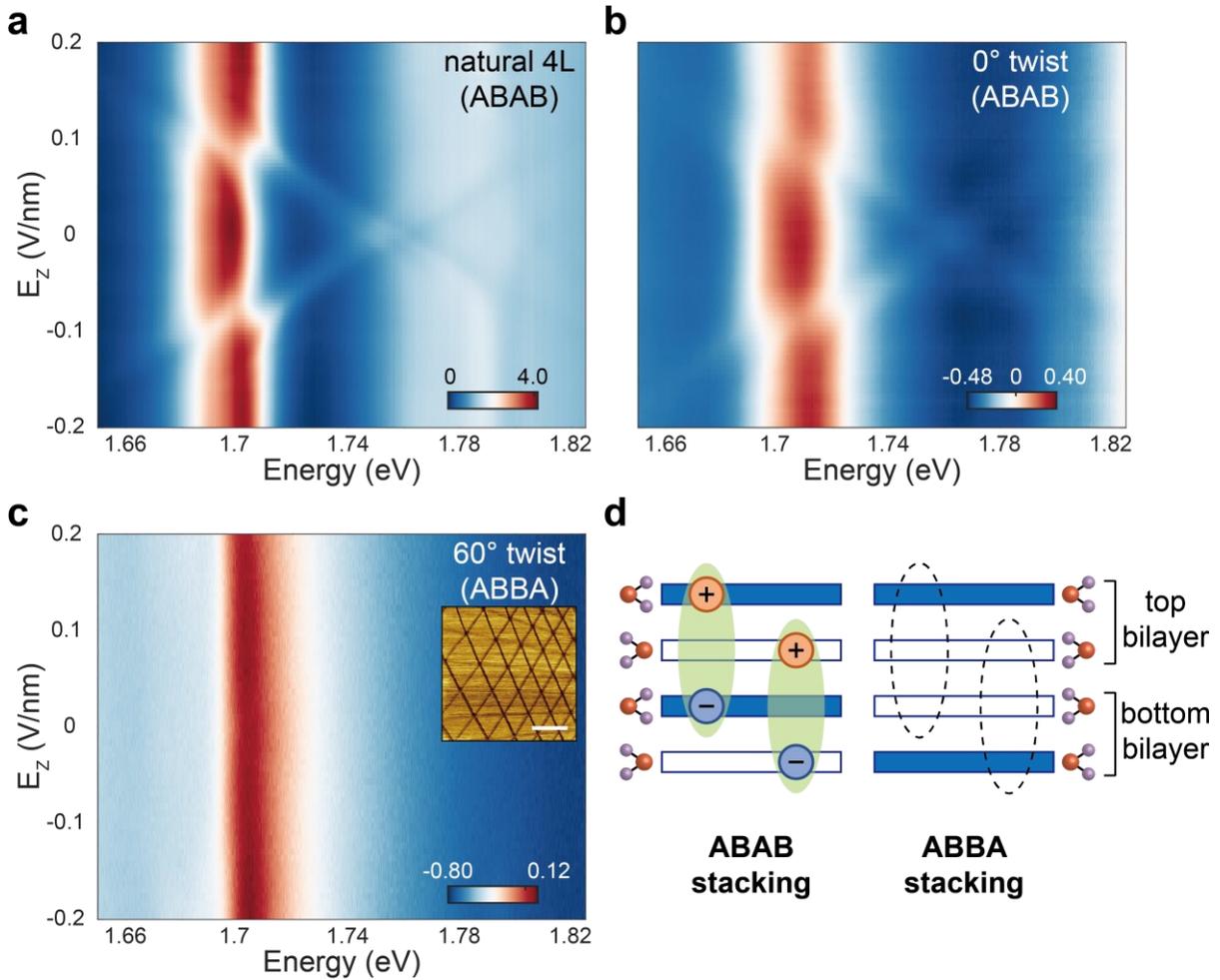

**Figure 3 | Stacking dependent every-other-layer exciton. a-c,** Electric field dependence of dR/R from natural 4-layer (**a**), near-0° twisted double bilayer WSe$_2$ (**b**) and near-60° twisted double bilayer WSe$_2$ (**c**). Inset: Triangular domains appear in the piezoresponse force microscopy image of a near-60° twisted double bilayer WSe$_2$ device. **d,** Cartoon of every-other-layer dipolar exciton formation in double bilayers with ABAB stacking, in comparison with the ABBA stacking.

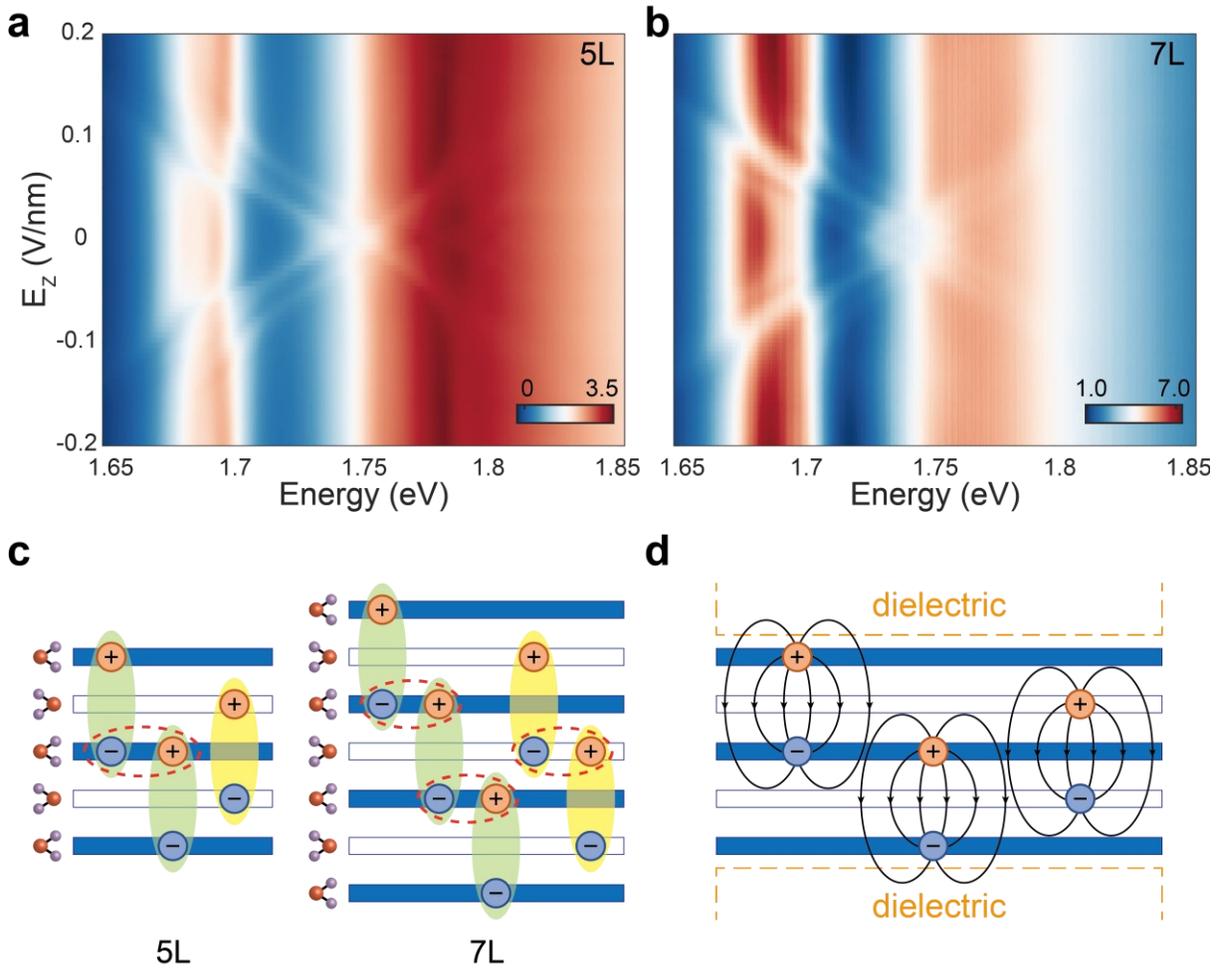

**Figure 4 | Layer dependence and multiple anti-crossings. a, b,** Multiple anti-crossings appear in the electric field dependence of dR/R from 5-layer (**a**) and 7-layer (**b**) 2H-WSe$_2$. **c,** Hopping of DXs in the out-of-plane direction through a virtual intralayer state in 5-layer and 7-layer. Dashed red ellipses stand for the intralayer exciton states which are coupled to two adjacent DXs. **d,** Illustration of different dielectric environment for DXs in a 5-layer geometry.

# Extended Data for

# Every-other-layer Dipolar Excitons in a Spin-Valley locked Superlattice

Yinong Zhang[1], Chengxin Xiao[2,3], Dmitry Ovchinnikov[1], Jiayi Zhu[1], Xi Wang[1,4], Takashi Taniguchi[5], Kenji Watanabe[6], Jiaqiang Yan[7], Wang Yao[2,3,*], and Xiaodong Xu[1,8,*]

Correspondence to: wangyao@hku.hk; xuxd@uw.edu

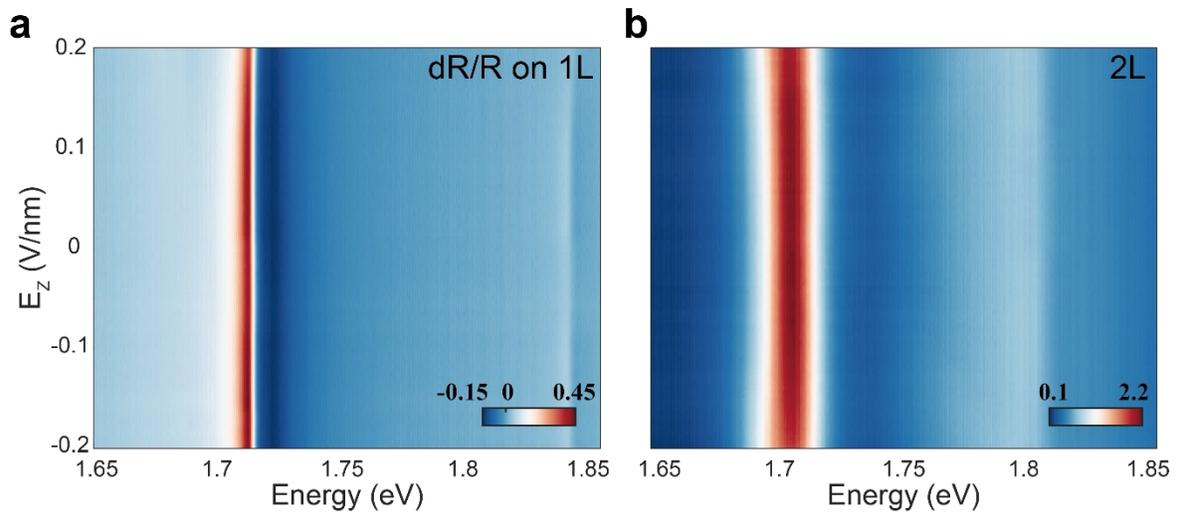

**Extended Data Fig. 1 | Optical reflectance spectra in monolayer and bilayer WSe$_2$. a-b,** Electric field ($E_z$) dependence of differential optical reflectance measurement in monolayer (a) and bilayer (b) WSe$_2$. Doping is fixed at zero. As expected, the every-other-layer dipolar exciton as well as the anti-crossing feature are not observed.

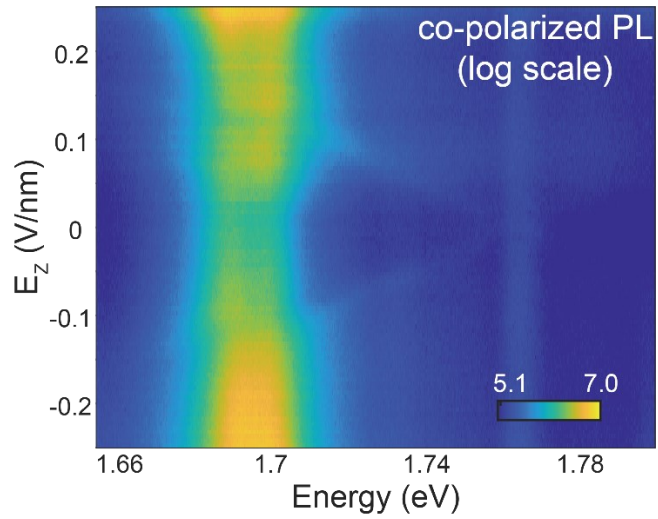

**Extended Data Fig. S2 | Electric field $E_Z$ dependence of photoluminescence spectra in trilayer WSe$_2$.** In addition to the dR/R data presented in the maintext, we examine photoluminescence (PL) spectra and its $E_Z$ field dependence on the same trilayer WSe$_2$. Near K-K direct transition region (around 1.7 eV), we find similar anti-crossing feature as dR/R spectra, supporting the formation of every-other-layer dipolar excitons. The field-independent PL feature around 1.76 eV is instrument artifact.

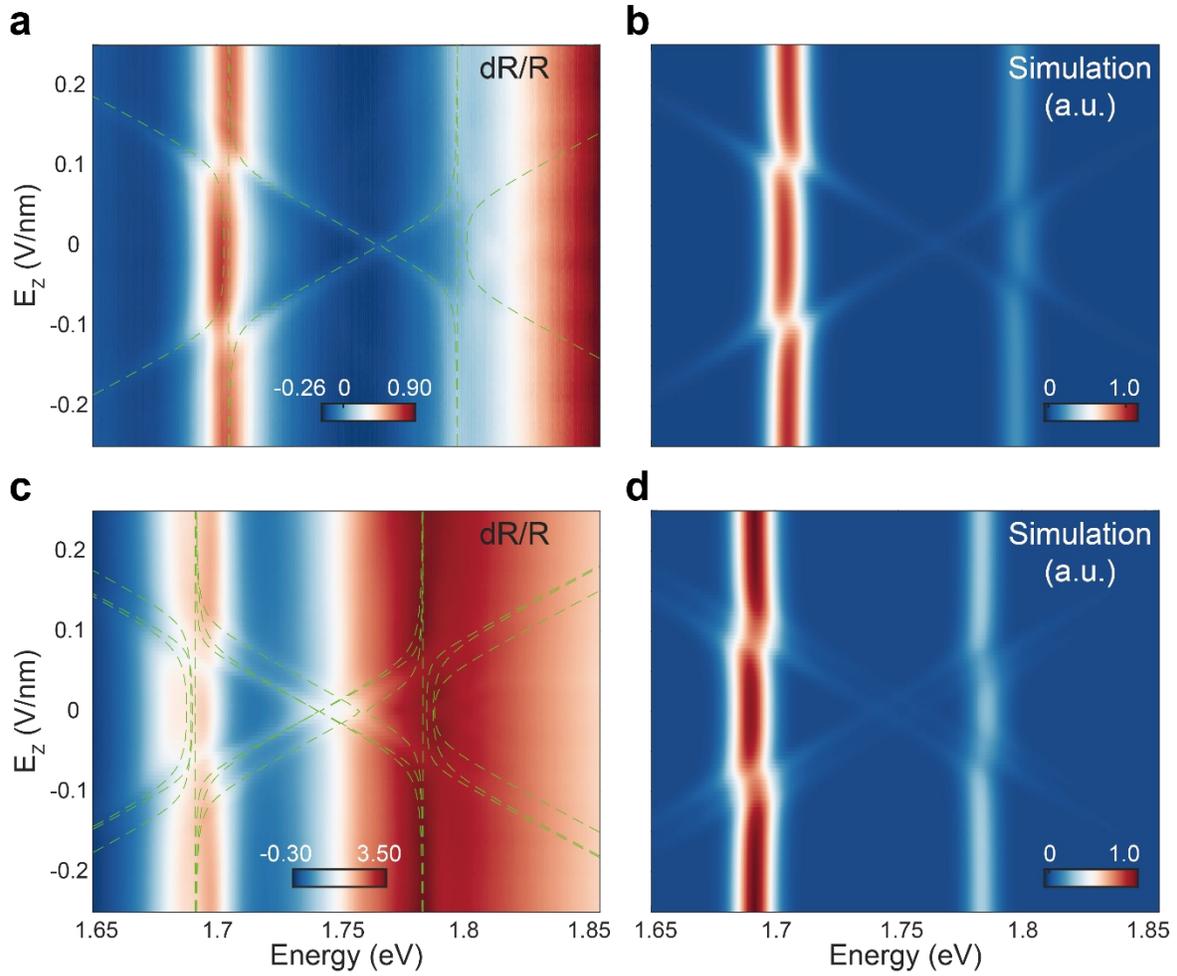

**Extended Data Figure 3 | Modeling and simulating the every-other-layer dipolar exciton. a,** $E_z$ dependence of measured differential reflectance spectra in trilayer $WSe_2$. The green dashed lines show the curve fitting results based on the Hamiltonian Eq. (1a) + (1b). **b,** Simulation results after considering the spectral intensity and width. **c, d,** Same plots for 5-layer $WSe_2$.